\documentclass{elsart}
\usepackage{graphicx,amssymb}
\journal{Physics Letters B}
\begin{document}
\begin{frontmatter}

%%%
%\begin{document}

%\begin{frontmatter}
\title{Gauging $U(1)$ symmetries and the number of right-handed neutrinos }

\author{J. C. Montero\thanksref{jcm},}
%\address{Instituto de F\'\i sica Te\'orica,
%Universidade Estadual Paulista\\ Rua Pamplona 145,
%01405-900 - S\~ao Paulo, SP, Brazil
%}
\ead{montero@ift.unesp.br}
\author{V. Pleitez\thanksref{vp}}
\address{Instituto de F\'\i sica Te\'orica,
Universidade Estadual Paulista\\ Rua Pamplona 145,
01405-900 - S\~ao Paulo, SP, Brazil
}
\ead{vicente@ift.unesp.br}
\thanks[jcm]{Partially supported by CNPq under the process 307807/2006-1}
\thanks[vp]{Partially supported by CNPq under the process
300613/2005-9}
%\ead[url]{authors.elsevier.com/locate/latex}

%\autho
%\address{Instituto de F\'\i sica Te\'orica,
%Universidade Estadual Paulista\\ Rua Pamplona, 145
%01405-900 - S\~ao Paulo, SP, Brazil
%}
%\ead{vicente@ift.unesp.br}
%%\ead[url]{authors.elsevier.com/locate/latex}

%\begin{frontmatter}
%\title{Gauging $U(1)$ symmetries and the number of sterile neutrinos
%}

\begin{abstract}
%The fact that the number of right-handed neutrinos that can be added to the
%minimal standard model is arbitrary can be seen as a suggestion that the
%symmetries of the interactions involving the known fermions should be larger.
%If these neutrinos are not sterile with respect to the new interactions,
%the anomaly cancellation can constraint their number.

In this letter we consider that assuming: a) that the only
left-handed neutral fermions are the active neutrinos, b) that $B-L$
is a gauge symmetry, and c) that the $L$ assignment is restricted to
the integer numbers, the anomaly cancellation imply that at least
three right-handed neutrinos must be added to the minimal
representation content of the electroweak standard model. However,
two types of models arise: i) the usual one where each of the three
identical right-handed neutrinos has total lepton number $L=1$; ii)
and the other one in which two of them carry $L=4$ while the third
one carries $L=-5$.
\end{abstract}
\begin{keyword} right-handed neutrinos, local $B-L$ symmetry, multi-Higgs models.
%\file{elsart}, document class, instructions for use
\PACS 14.60.St; 11.30.Fs; 12.60.Fr

\end{keyword}
\end{frontmatter}

%%%%%%%%%%%%%%%%%555
It is well known that it is possible to enlarge the representation
content of the minimal electroweak standard model (ESM)
by adding an arbitrary number of right-handed neutrinos. Since they
are sterile under the interactions of that model they do not contribute to
the anomaly cancellation of the gauge symmetries, then nothing determine
their number. Until now, it has been a question of taste to consider a
particular number of these fields in extensions of the model. It is also
well known that within the ESM (no right-handed neutrinos) both, baryon
($B$) and total lepton ($L$) numbers, are conserved automatically up to
anomaly effects: both global $U(1)_B$ and $U(1)_L$ are
anomalous~\cite{thooft} (but their consequences are well suppressed
at least at zero temperature) and only the combination $U(1)_{_{B-L}}$ is a global
anomaly free symmetry if right-handed neutrinos are added for cancelling the mixed
gauge-gravitational anomaly~\cite{eguchi}. When  the $B-L$ symmetry is gauged, it becomes anomaly
free but only again if an appropriate number of right-handed neutrinos is added, but this time
they also must cancel out other anomalies, like the cubic one, induced by the active left-handed
neutrinos. For instance, adding one per generation solve again the problem.

In this Letter we will propose extensions of the standard model in which $B-L$
appears as a local symmetry. Many of the extension of the SM in which $B-L$ is a gauge symmetry
are based on $SMG\otimes U(1)_{B-L}$ gauge symmetry~\cite{ux}. However, in those models, since
$SMG$ is the gauge symmetry of the SM, the usual Higgs doublet does not carry the $U(1)_X$ charge,
and then the electric charge $Q$ is given in terms of the $SU(2)_L$ and $U(1)_Y$
generators alone. This implies important phenomenological differences with the models
that we will consider below, in which the electric charge includes the $U(1)$ extra
generators. Other models with extra $U(1)$ factors are based on grand unified
scenarios~\cite{e6,lrgut}. There are also models with an extra $U(1)$ factor and a
$Z^\prime$ with non-universal couplings to fermions in which right-handed interactions
single out the third generation~\cite{valencia}. The difference between models with
additional $U(1)$ groups not inspired in unified theories is that the neutral current
parameters in the latter case must satisfied some relations~\cite{barr} that do not
exist in the former. For this reason these parameters are more arbitrary in our models
than in models like those in Refs.~\cite{e6,lrgut}. In these sort of model there is  $Z-Z_X$
mixing in the mass matrix at the tree level. Of course, mixing in the kinetic term is
possible~\cite{holdom}, but we will assume that we are working in a basis in which the
kinetic mixing vanishes. For a review of the phenomenology of the extra neutral vector
boson see Ref.~\cite{zprime}.

Hence, we will consider an extension of the $SMG$ based on the following gauge symmetry:
\begin{eqnarray}
SU(3)_C\otimes SU(2)_L\otimes U(1)_{Y^\prime}\otimes U(1)_{B-L}
\nonumber \\ \downarrow  \langle\phi\rangle \nonumber \\
SU(3)_C\otimes SU(2)_L\otimes U(1)_Y \nonumber \\ \downarrow \langle\Phi\rangle\nonumber
\\ SU(3)_C\otimes U(1)_{em},
\label{group}
\end{eqnarray}
where $Y^\prime$ is chosen to
obtain the hypercharge $Y$ of the standard model, given by
$Y=~Y^\prime+~(B-L)$. Thus, in this case, the charge operator is given by
\begin{equation}
    \frac{Q}{e}=I_3+\frac{1}{2}\,\left[Y^\prime + (B-L)\right].
    \label{gn}
\end{equation}

The simplest possibility is adding three right-handed neutrinos with the same lepton number
as that of the left-handed ones. In this case $B-L$ is anomaly free. We also add a complex neutral
scalar $\varphi$ that because of  $\langle\varphi\rangle\not=0$, breaks the $U(1)_{_{B-L}}$ gauge symmetry.
The quantum number of the fields in this model are shown in Table~\ref{table1}.

\begin{table}
\begin{eqnarray*}
\begin{array}{|c||c|c|c|c|c|c|}
  \hline
 \phantom{u_L} & I_3 & I & Q &  Y^\prime & B-L & Y\\ \hline\hline
  \nu_{eL} & 1/2 & 1/2 & 0  & 0 & -1 &-1\\ \hline
  e_L     & -1/2 & 1/2 & -1  & 0 & -1 &-1\\ \hline
  e_R     & 0 & 0 & -1 & -1 & -1 &-2\\ \hline
 n_R & 0 & 0 & 0  & 1 & -1 & 0 \\ \hline
  u_L& 1/2 & 1/2 & 2/3  & 0 & 1/3 & 1/3\\ \hline
  d_L & -1/2 & 1/2 & -1/3  &0 & 1/3 &1/3\\ \hline
  u_R & 0 & 0 & 2/3  &1 & 1/3  & 4/3\\ \hline
  d_R & 0 & 0 & -1/3  &-1 & 1/3 &-2/3 \\ \hline
  \varphi^+ & 1/2 & 1/2 & 1  &1 & 0 & 1\\ \hline
  \varphi^0 & -1/2 & 1/2 & 0  &1 & 0 & 1\\ \hline
  \phi & 0 & 0 & 0 &  -2 & 2 &0\\ \hline
\end{array}
\end{eqnarray*}
\caption{Quantum number assignment in the model with three
identical right-handed neutrinos.}
\label{table1}
\end{table}

The model has three real neutral gauge bosons $W^3$, $\mathcal{A}$,
$\mathcal{B}$ corresponding to the $SU(2)_L$, $U(1)_{Y^\prime}$, and
$U(1)_{B-L}$ factors respectively, are mixtures of the photon, $A$,
and two massive neutral bosons, $Z_1\approx Z$, and $Z_2\approx
Z^\prime$, fields. The model introduces deviations of the $\rho$
parameter, at the tree level, that can be parameterized by the $T$
parameter defined, in absence of new charged $W$-like vector bosons,
and neglecting the contributions of the Majorana neutrinos which
contributions to the $T$-parameter may have either sign, as
$\hat{\alpha}(M_Z)T\equiv-\Pi^{new}_{ZZ}(0)/M^2_{Z_1}$, where
$\Pi^{new}_{ZZ}(0)=M^2_{Z_1}-(g^2v^2/4c^2_W)$, being $M^2_{Z_1}$ the
exact mass of the lighter neutral vector boson that we are not
showing here. We obtain $\Delta\rho~=~\hat{\alpha}(M_Z) T\approx
(g^{\prime\,4}/4)\,\bar{v}^2$. This implies in the lower bound
$u~>~(10^4\,
g^{\prime\,2})\,\textrm{GeV}\,~>~4\pi\,(10^4\alpha^2s^2_W/c^4_W)$
GeV, in order to be consistent with the experimental
data~\cite{pdg}. The scalar singlet contributes less to the mass of
the lighter vector boson as its VEV is higher, i.~e., if
$u\to\infty$ then $Z_1\to Z$ and $Z_2$ decouples. Besides, since we
are working in a basis where there is no kinetic mixing between the
$U(1)_{Y^\prime}$ and $U(1)_{_{B-L}}$ gauge bosons, there are no
tree level contributions to the $S$ and $U$ parameters~\cite{appelquist}.

Quark and charged lepton Yukawa interactions are the same as in the
ESM. However, the neutrino mass terms are Dirac terms involving the
left-handed leptons $\Psi=(\nu_l\, l)^T$, and the scalar doublet
$\Phi$, $\overline{\Psi}_{aL}\,G^D_{a\alpha}\,\Phi\, n_{\alpha R}$,
and Majorana terms involving the singlet $\phi$,
$\phi\,\overline{(n_{a R})^c}\,G^M_{ab}\,n_{b R}$, where
$a=e,\mu,\tau$ and we have omitted summation symbols. If $\langle
\Phi\rangle=v/\sqrt{2}\simeq 174$ GeV the neutrino Dirac masses are
of the same order of magnitude (up a fine tuning in $G^D$). Hence,
in this case for implementing the seesaw mechanism we have to have
that $\langle\phi\rangle=u/\sqrt{2}\gg \langle\Phi\rangle$ and there is no
natural possibility for having light right-handed neutrinos.
However, if the doublet $\Phi$ is different from the doublet which
gives masses for quarks and charged leptons, $\langle\Phi\rangle$
can be smaller than the electroweak scale, and $\langle\phi\rangle$
is not necessarily a large energy scale and could be constrained
only by the phenomenological allowed value for the $Z^\prime$ mass.
More details of the phenomenology of this model at LHC and ILC
energies and its comparison with other models with a $Z^\prime$ will
be given elsewhere~\cite{elaine}.

One condition for having $B-L$ as a local anomaly free symmetry is that considered above.
The number of right-handed neutrinos is $N_R=3$, one per generation, and all of them  carry
$Y^\prime(n_{\alpha R})= -(B-L)(n_{\alpha R})=-1,\forall \alpha$.
However, it is possible to consider these quantum numbers as free parameters. In this
case, in order to generate Dirac mass for neutrinos, it is necessary to introduce scalar 
doublets that carry also $Y^\prime$ and $B-L$ charges.
The quantum numbers of the new fields are shown in Table~\ref{table2}. Since the number
of right-handed neutrinos and their $B-~L$ assignment are free parameters, the only
constraint is that they have to cancel the cubic and linear anomalies  of the three
active left-handed neutrinos altogether (not generation by generation) by having the
appropriate $B-L$ attribution which is not necessarily an integer number.

The right-handed neutrinos contribute to the following anomalies:
\begin{equation}
\textrm{Tr}\,[U(1)_{_{B-L}}]^2U(1)_{Y^\prime},\;
\textrm{Tr}\,[U(1)_{Y^\prime}]^2U(1)_{_{B-L}},\;
\textrm{Tr}[U(1)_{Y^\prime}]^3, \;\textrm{Tr}[U(1)_{_{B-L}}]^3,
\label{ano}
\end{equation}
that imply the following equations:
\begin{eqnarray}
&& \sum_{\alpha=1}^{N_R}
Y^\prime(n_{\alpha R}) (B-L)^2(n_{\alpha R})=3,\quad
 \sum_{\alpha=1}^{N_R} Y^{\prime\,2}(n_{\alpha R}) (B-L)(n_{\alpha R})=-3,
\nonumber \\ &&
\sum_{\alpha=1}^{N_R}Y^{\prime\,3}(n_{\alpha R})=3,\qquad\qquad\qquad
\sum_{\alpha=1}^{N_R}(B-L)^3(n_{\alpha R})=-3,
\label{e1}
\end{eqnarray}
besides the two conditions for cancelling the gauge--gravitational anomaly:
\begin{equation}
\sum^{N_R}_{\alpha=1}\,Y^\prime(n_{\alpha R})=3,\;
\sum_{\alpha=1}^{N_R}(B-~L)(n_{\alpha R})=-3.
\label{gg}
\end{equation}

However, the condition
$[Y^\prime+(B-L)](n_{\alpha R})=0$, for $\alpha$ fixed, has to be imposed
in order to have right-handed neutrinos that are sterile with respect to the standard model
interactions, so that the anomaly cancellation conditions in Eqs.~(\ref{e1}) and (\ref{gg})
are reduced to the following equations:
\begin{equation}
\sum_{\alpha=1}^{N_R}Y^{\prime\,3}(n_{\alpha R})=3,\quad
\sum_{\alpha=1}^{N_R}\,Y^\prime(n_{\alpha R})=3.
\label{e3}
\end{equation}

\begin{table}
\begin{eqnarray*}
\begin{array}{|c||c|c|c|c|c|c|}\hline
\phantom{u_L} & I_3 & I & Q &  Y^\prime & B-L & Y\\ \hline\hline
n_{1R} & 0 & 0 & 0  & Y^\prime_1 & -Y^\prime_1 & 0 \\ \hline
n_{2R} & 0 & 0 & 0  & Y^\prime_2 & -Y^\prime_2 & 0 \\ \hline
n_{3R} & 0 & 0 & 0  & Y^\prime_3 & -Y^\prime_3 & 0 \\ \hline
\varphi^0_i & 1/2 & 1/2 & 1 & Y^\prime_i  & -Y^\prime_i-1  & -1\\ \hline
  \varphi^-_i & -1/2 & 1/2 & 0  & Y^\prime_i & -Y^\prime_i-1 & -1\\ \hline
  \phi_s & 0 & 0 & 0 &  Y^\prime_s & -Y^\prime_s &0\\ \hline
\end{array}
\end{eqnarray*}
\caption{Quantum number assignment in the model with three
non-identical right-handed neutrinos. The number of doublet and singlet
scalars depend on the values for $Y^\prime_{1,2,3}$. The other fields have the
quantum number given in Table~\ref{table1}.}
\label{table2}
\end{table}

In solving Eqs.~(\ref{e3}), we will also assume that there is no vectorial neutral
leptons, i.e., $Y^\prime(N_{1L})=Y^\prime(N_{1R})$, and also that no neutral mirror leptons,
i.e., $Y^\prime(N_{1R})=-Y^\prime(N_{2R})$, are added. For Majorana fermions both cases are 
equivalent since $N_{1L}$ is related by CP to its right-handed conjugate.
It means that having found a solution for the Eqs.~(\ref{e3}), no extra terms vanishing among 
themselves are introduced: these sort of leptons would only cancel out their own anomalies, not 
the anomalies induced by the active left-handed neutrinos. They just add ``0'' to the left side 
of Eqs.~(\ref{e3}) and, hence, are meaningless to our strategy.

Solving the constraint equations in Eq.~(\ref{e3}), we have found that when $N_R=1$ they have
no solutions; when $N_R=2$, there are only complex solutions. In the case of $N_R=3$, we can
only find two $Y^\prime$ in terms of the third one, say, $Y^\prime(n_{1R})\equiv Y^\prime_1$
and $Y^\prime(n_{2R})\equiv Y^\prime_2$ in terms of $Y^\prime(n_{3R})\equiv Y^\prime_3$, and
the solutions are:
\begin{equation}
2Y^\prime_1= 3-Y^\prime_3\pm \frac{1-Y^\prime_3 }{ Y^\prime_3-3 }\,
R(Y^\prime_3),\;\; 2Y^\prime_2= 3-Y^\prime_3\mp \frac{1-Y^\prime_3 }
{ Y^\prime_3-3 }\, R(Y^\prime_3),
\label{nova1}
\end{equation}
where, $R(x)=[(x-3)(x+5)]^{1/2}$.

From the last equations we obtain again the solution with identical right-handed neutrinos,
i.e., all of them carrying $Y^\prime_1=Y^\prime_2=Y^\prime_3\equiv Y^\prime=1$ and
$(B-L)_1=(B-L)_2=(B-L)_3\equiv B-L=-1$, we have already studied above.
However, there is also other solution concerning only integer values of $Y^\prime$
and $B-L$ (we recall that these numbers are integer for
charged leptons and active neutrinos): two right-handed neutrinos with, say, $Y^\prime_1=
Y^\prime_2=-(B-L)_1=-(B-L)_2=4$ and the third one with $Y^\prime_3=-(B-L)_3=-5$. There are
also real non-integer solutions but we will not consider them here.
For $N_R=4$ we have also found an infinite number of
real (non-integer) solutions for the assignment of $Y^\prime=-(B-L)$ for the right-handed
neutrinos, that we are not showing explicitly. The only integer solutions are those of the
$N_R=3$ but with the fourth neutrino carrying $Y^\prime=0$. However we are not considering
right-handed neutrinos which are singlets of the new interactions. We have also worked the
cases for $N_R=5,6$ and found out that there are several solutions with $Y^\prime$ integer.
For instance, $Y^\prime_i=(-11,-2,-1,7,10)$ for $N_R=5$; and $Y^\prime_i=(-6,-6,1,3,4,7)$
for $N_R=6$. In general for $N_R\geq 5$ it is possible that there exist an infinite set of
solutions. Hence, only the case $N_R=3$ has just two solutions of this sort: $Y^\prime=(1,1,1)$,
which is the usual one, and the exotic $(-5,4,4)$ one. We will consider below a model based on
the exotic solution for the case of three right-handed neutrinos.

In this model the analysis of the $T$ parameter is more complicated than in the first model
because, besides the Majorana neutrinos, there are additional
Higgs doublets which, unlike the Dirac fermion case which are always positive, give
contributions to the $T$-parameter with either sign~\cite{majorana,dubletos}. We will shown
these explicitly elsewhere. Here, we will give details only of the scalar and the Yukawa sectors.

The scalar sector of the theory is constituted by several doublets and singlets.
For instance, the scalar sector which interacts in the lepton sector could be: the usual
doublet with $Y=+1$, here denoted by $\Phi_{_{SM}}$, two doublets with $Y=-1$: one,
denoted by $\Phi_1$, with $Y^\prime=-4$, and $(B-L)=+3$, and the other, $\Phi_2$, with
$Y^\prime=5$, and $(B-L)=-6$; and three complex scalar singlets ($Y=0$): $\phi_1$ with
$Y^\prime=-(B-L)=-8$, $\phi_2$ with $Y^\prime=-(B-L)=10$, and $\phi_3$ with
$Y^\prime=-(B-L)=1$. Notice that whenever the scalar doublets carry a non-zero $B-L$, it
means that these doublets contribute to the spontaneous violation of this number, which is
also induced by the complex scalar singlets.

This model is interesting for introducing three scales for the Majorana masses of the right-handed neutrinos.
With these fields and the leptons we have the Yukawa interactions
(omitting summation symbols)
\begin{eqnarray}
-\mathcal{L}^\nu_{\textrm{yukawa}}&=&
\overline{\Psi}_{aL}G^D_{a m}\Phi_1 n_{m R}+\overline{\Psi}_{aL}G^D_{a3}
\Phi_2n_{3 R}+
\phi_1\,\overline{(n_{m R})^c}\,G^M_{mn}\,n_{n R} \nonumber \\ &+&
\phi_2\,\overline{(n_{3 R})^c}\,G^M_{33}n_{3 R}+ \phi_3
\overline{(n_{m R})^c}\,G^M_{m3}\,n_{3 R}+H.c.,
\label{see2}
\end{eqnarray}
where $m,n=1,2$. 

Not all of the Majorana mass terms, for the right-handed neutrinos, are necessarily
too large since only one of the singlets has to have a large VEV so that the breaking of
the $B-L$ symmetry occurs at a high energy scale. In fact, two of them can be light enough
to implement the $3+2$ neutrino scheme, with $CP$ violation, as in Ref.~\cite{maltoni}.
If some singlet neutrinos are heavy but not too much, effects of them could be detectable
at the LHC~\cite{lhc}, linear~\cite{ilc} or $e$-$\gamma$~\cite{egamma} colliders, or in low
 energy processes~\cite{tau2}. In particular lepton colliders would be appropriate for
discovering these sort of neutrinos~\cite{ll}. If $n_{\alpha R}$ are heavier than all
the physical scalar fields which are almost doublets, the decays $n_{\alpha R}\to l^\pm
h^\mp$ are kinematically allowed, and hence $h^\pm \to h^0+W^{\pm *}$ or $h^\pm\to
\bar{q}q^\prime$, where $h^+(h^0)$ denotes any charged (neutral) physical scalar,
$q,q^\prime$ are quarks with different electric charge, and $W^{\pm *}$ is a virtual
vector boson. Hence, in this model, only the lightest of the neutral almost scalar
singlets would be a candidate for dark matter~\cite{darkmatter}.

In the model with quantum number given in Table~\ref{table1}, the more general 
$SU(2)_L\otimes U(1)_{Y^\prime}\otimes U(1)_{B-L}$ invariant
scalar potential for the doublet $\Phi$ and the singlet $\phi$, is given by

\begin{equation}\label{vip}
    V(\Phi,\phi)=\mu^2_1 \vert\Phi\vert^2 + \mu^2_2 \vert\phi\vert^2 +
    \lambda_1 \vert\Phi^\dag \Phi\vert^2 + \lambda_2 \vert\phi^\dag 
    \phi\vert^2+
    \lambda_3 \vert\Phi\vert^2\vert\phi\vert^2.
\end{equation}
Doing as usual the shifted as
$\varphi^0~\!\!\!=\!\!\!~\frac{1}{\sqrt2}(v~+~H~+~iF)$ and $\phi~=~
\frac{1}{\sqrt2}(u~+~S~+~iG)$, so
that the constraint equations are given by:

\begin{equation}\label{vinculos}
v\left(  \mu_1^2+\lambda_1v^2+\frac{\lambda_3}{2}u^2\right) = 0,\;
u\left(  \mu_2^2+\lambda_2u^2+\frac{\lambda_3}{2}v^2\right) = 0.
\end{equation}
We will choose real $v,u\not=0$ solutions for simplicity. 
We also must have $\lambda_1,\lambda_2>0$, in order to the scalar
potential be bounded from below, and $\lambda_3^2<4\lambda_1
\lambda_2$, to assure we have a minimum.
The mass square matrix in the
basis $(H,S)$, after the use of Eq.(\ref{vinculos}), is given by
\begin{equation}\label{ms}
M^2_S\,=\,\left(
  \begin{array}{cc}
    2\lambda_1 v^2 & \lambda_3uv \\
    \lambda_3uv & 2\lambda_2 u^2 \\
  \end{array}
\right)\,,
\end{equation}
with $\textrm{Det}\,M^2_S\neq 0$ by the above conditions. The exact eigenvalues 
for the mass square matrix are:
\begin{equation}
m^2_{1,2}= \lambda_1 v^2+\lambda_2 u^2 \pm
\left[\left(\lambda_1 v^2+\lambda_2
u^2\right)^2\!\!\!\!-\!\!\left(4\lambda_1\lambda_2\!\!-\!\!
\lambda_3^2\right)u^2v^2\right]^{\frac{1}{2}},
\label{mass12}
\end{equation}
which can be approximate by considering $u\gg v$ (but still
arbitrary),
\begin{equation}\label{maprox}
m_1^2\approx2\lambda_1\left(1-\frac{\lambda_3^2}{4\lambda_2\lambda_1}\right)
\,v^2\,,\quad
m_2^2\approx2\lambda_2u^2+\frac{\lambda_3^2}{2\lambda_2}\,v^2.
\end{equation}

Notice that the heavier neutral boson has a mass square proportional to $u^2$,
$m_2>m_1$. The exact eigenvectors are give by

\begin{equation}\label{eigenvectorsscalars}
H_1= -\frac{1}{\sqrt{N_1}}\left(\frac{a-\sqrt{a^2+b^2}}{b}\;H + S\right),\;
H_2= \frac{1}{\sqrt{N_2}}\left(\frac{a+\sqrt{a^2+b^2}}{b}\;H + S\right),
\end{equation}
where $a=\lambda_1v^2-\lambda_2u^2$, $b=\lambda_3uv$, and $N_{1,2}=1+(\sqrt{a^2+b^2}\mp a)^2/b^2$.
We have maximal mixing when $\lambda_1/\lambda_2=u^2/v^2$. 
The eigenvectors in Eq.~(\ref{eigenvectorsscalars}) can be written as follows
\begin{equation}
\left( \begin{array}{c}
H_1 \\ H_2\end{array}\right)=\left( \begin{array}{cc}
\cos\theta & \sin\theta \\
-\sin\theta &\cos\theta\end{array}\right) \left( \begin{array}{c} 
H\\ S\end{array}\right).
\label{oba}
\end{equation}
This implies a reduction on the value of the couplings of the Higgs to standard model particles, 
$h_1=h\cos\theta$, and $h_2=h\sin\theta$, where $h$ denotes any of the SM coupling
constants for the Higgs scalar. Depending on the value of the angle $\theta$ we 
can suppress the Higgs decays making the SM Higgs invisible even at the LHC. This 
effect has been considered in literature when the added scalar singlet is real~\cite{bij}.
The  would be Goldstone boson, $F$ and $G$ in the unitary gauge, are absorbed by the longitudinal 
components of $Z$ and $Z^\prime$ respectively.

On the other hand, for the second model the most general $SU(2)_L\otimes U(1)_{Y^\prime}\otimes 
U(1)_{_{B-L}}$ invariant potential may be written as
\begin{eqnarray}
V_{_{B-L}}&=&V_{_{SM}}(\Phi_{_{SM}})+
\mu^2_{11}\Phi^\dagger_1\Phi_1+  \mu^2_{22}\Phi^\dagger_2\Phi_2+
\lambda_1 \vert\Phi^\dagger_1\Phi_1\vert^2 +
\lambda_2 \vert\Phi^\dagger_2\Phi_2\vert^2+
\lambda_3\vert\Phi_1\vert^2\vert\Phi_2\vert^2 \nonumber \\
&+&\lambda_4 (\Phi^\dagger_1\Phi_2)(\Phi^\dagger_2\Phi_1)+
\lambda_{_{SMi}}\vert\Phi_{_{SM}}\vert^2\vert\Phi_i\vert^2
+\lambda^\prime_{_{SM\alpha}}
\vert \Phi_{_{SM}}\vert^2\vert\phi_\alpha\vert^2+
\lambda^\prime_{i\alpha}
\vert\Phi_i\vert^2\vert\phi_\alpha\vert^2 \nonumber \\ &+&
\mu^{2}_\alpha\vert\phi_\alpha\vert^2 +
\lambda^\prime_\alpha \vert\phi^*_\alpha\phi_\alpha\vert^2+
[\Phi^\dagger_1\Phi_2(\kappa\,\phi_1\phi^*_3+\kappa^\prime \phi^*_2\phi_3)+
\lambda^{\prime\prime}(\phi^*_3)^2\phi_1\phi_2+H.c.]\nonumber \\ &+&
\lambda_{\alpha\beta}(\phi^*_\alpha\phi_\alpha)(\phi^*_\beta\phi_\beta),
\label{potential}
\end{eqnarray}
where $i,j=1,2$ and $\alpha=1,2,3$ (we have omitted summation symbols), in the last term $\alpha<\beta$;
and since $\Phi_{_{SM}}$ is the usual Higgs doublet of the SM,  $V_{_{SM}}(\Phi_{_{SM}})$
denotes the respective potential. 

The constraint equations coming from the linear terms of the scalar potential in 
Eqs.~(\ref{potential}) are:
\begin{eqnarray}
&&v_1[2\mu^2_{11}+2\lambda_1v^2_1+(\lambda_3+
\lambda_4)v^2_2
+\lambda_{_{SM1}}v^2_{_{SM}} +\lambda^\prime_{11}v^2_{s_1}+
\lambda^\prime_{12}v^2_{s_2}+\lambda^\prime_{13}v^2_{s_3}]\nonumber \\&&
+v_2(\kappa v_{s_1}v_{s_3}+
\kappa^\prime v_{s_2}v_{s_3})=0,
\nonumber \\ &&
v_2[2\mu^2_{22}+2\lambda_2v^2_2+(\lambda_3+
\lambda_4)v^2_1
+\lambda_{_{SM2}}v^2_{_{SM}}+\lambda^\prime_{21}v^2_{s_1} +
\lambda^\prime_{22}v^2_{s_2}+\lambda^\prime_{23}v^2_{s_3}]
\nonumber \\ &&+v_1(\kappa v_{s_1}v_{s_3}
+\kappa^\prime v_{s_2}v_{s_3})=0,
\nonumber \\ &&
v_{_{SM}}[2\mu^2_{_{SM}}+\lambda_{_{SM1}}v^2_1
+2\lambda_{_{SM}}v^2_{_{SM}}+\lambda_{_{SM2}}v^2_{2} + 
\lambda^\prime_{_{SM1}}v^2_{s_1}+\lambda^\prime_{_{SM2}}v^2_{s_2}+ 
\lambda^\prime_{_{SM3}}v^2_{s_3}]=0,
\nonumber \\&&
v_{s_1}[2\mu^{2}_1 +2\lambda^\prime_1 v^2_{s_1}+\lambda^\prime_{_{SM1}}v^2_{_{SM}}+\lambda_{12}v^2_{s_2}
+\lambda_{13}v^2_{s_3}
+\lambda^\prime_{11}v^2_1 +\lambda^\prime_{21}v^2_2+\lambda_{12}v^2_{s_2}]
\nonumber \\&& \lambda^{\prime\prime}v_{s_2}v^2_{s_3}+\kappa v_1v_2v_{s_3}=0,
\nonumber \\&&
v_{s_2}[2\mu^{2}_2 +2\lambda^\prime_2 v^2_{s2}+
\lambda^\prime_{_{SM2}}v^2_{_{SM}}+
\lambda_{12}v^2_{s_1}
+\lambda_{23}v^2_{s_3}+
\lambda^\prime_{12}v^2_1+\lambda^\prime_{22}v^2_2]\nonumber \\ &&+
\lambda^{\prime\prime}v_{s_1}v^2_{s_3}+\kappa^\prime v_1v_2v_{s_3}
=0,\nonumber \\&&
v_{s_3}[2\mu^{2}_3 +2\lambda^\prime_3 v^2_{s_3}+ 
\lambda^\prime_{_{SM3}}v^2_{_{SM}}+\lambda_{13}v^2_{s_1}
+\lambda_{23}v^2_{s_2}
+\lambda^\prime_{13}v^2_1+\lambda^\prime_{23}v^2_2]
\nonumber \\&&+2\lambda^{\prime\prime}v_{s_1}v_{s_2}+\kappa v_1v_2v_{s_1}+\kappa^\prime v_1v_2v_{s2} =0,
\label{con1}
\end{eqnarray}
and we have also used the VEVs as being real for the sake of simplicity. With this potential if 
$\lambda^{\prime\prime},\kappa,\kappa^\prime\not=0$ all
VEVs have to be different from zero and it is possible to give to all fermions masses with the correct
values. This model has extra global $U(1)$ symmetries as can be verified by the number of neutral Goldstone bosons: 
there are four of them. Notice that only the fields carrying exotic values of $Y^\prime$ and $B-L$  can carry the
charge of the extra global symmetries. Hence, these extra symmetries are restricted to the exotic scalars 
and neutrino singlets, and from Eqs.~(\ref{see2}), we have the following equations:
\begin{eqnarray}
&&\zeta(\Phi_1)+\zeta(n_{mR})=0,\;\zeta(\Phi_2)+\zeta(n_{3R})=0,\;\zeta(\phi_1)+2\zeta(n_{mR})=0,\nonumber \\&&
\zeta(\phi_2)+2\zeta(n_{3R})=0,\;\zeta(\phi_3)+\zeta(n_{mR})+\zeta(n_{3R})=0,
\label{extra}
\end{eqnarray}
where $\zeta(f)$ denotes the $U(1)_\zeta$ charge of the field $f$. Fermionic left-handed doublets, electrically charged 
right-handed singlets and the scalar doublet $\Phi_{_{SM}}$ do not carry this sort of new charges. There are two solutions 
for the equations above that we will denote $\zeta=X,X^\prime$: i) $X(\Phi_1)=-X(n_{mR})=1,X(\Phi_2)=-X(n_{3R})=1$, $X(\phi_1)=
X(\phi_2)=X(\phi_3)=2$; and ii) $2X^\prime(\Phi_2)=X^\prime(\phi_2)=2X^\prime(\phi_3)=-2X^\prime(n_{3R})=-2$ and the other fields
no carrying this charge. It worth noting that extra Goldstone bosons arise in supersymmetric models with extra $U(1)$ 
factors and several scalar singlets under the SM gauge symmetries~\cite{langacker}. However, in the present model, 
this is not a flaw because the extra Goldstone bosons, denoted by $G_X$ and $G_{X^\prime}$, can be almost singlets:
$G_X$ can always be made almost singlet, $G_X\sim \phi_1$; $G_{X^\prime}$ 
may have its main projection on $\phi_2$ or $\phi_3$. Anyway, the extra Goldstone bosons are not a problem in this model also 
because they couple mainly to active and sterile neutrinos, hence its consequences may be important only on cosmological 
scales. In the scalar (CP even) sector all fields are massive.

Another possibility is to avoid the appearance of $G_X$ and $G_{X^\prime}$.
First, note that interactions that can break those symmetries are forbidden
by the $U(1)_{Y^\prime}$ and $U(1)_{_{B-L}}$ symmetries that in the present model are local symmetries.
Hence, it is not allowed to break directly and softly the global $U(1)_\zeta$ symmetries. One way to solve this 
issue is to add non-renormalizable operators that are invariant under the gauge symmetry of the model.
For instance $h\,(\phi_1^*\phi_1)(\phi_2^*\phi_2)(\phi_3^*\phi_3)/\Lambda^2$, where $\Lambda$ is an energy scale
higher than the electroweak scale, and $h$ is a dimensionless constant.  When the singlets get the VEVs
they induce terms like $\mu_{123}\phi_1\phi_2\phi_3$, where $\mu_{123}=hv^*_{s1}v^*_{s2}v^*_{s3}/\Lambda^2$.
When terms like that are introduced they modified the last three constraint equations in (\ref{con1}) and
the Goldstone bosons are reduced to just two: $G_X$ and $G_{X^\prime}$ have disappeared. 
Notice that $Y^\prime$ and $B-L$ are only hidden because the original dimension six operators are invariant under 
these symmetries.

It is interesting to note that the SM is anomalous with respect to the mixed global
$(B-L)$-gravitational anomaly. It is  canceled out if right-handed neutrinos are introduced. In
this case the condition for cancelling that anomaly, for the three generation case, is
$\sum_{\alpha=1}^{N_R} (B-L)(n_{\alpha R})=-3$. For instance, if $N_R=1$ the unique right-handed
neutrino must carry $L=3$; if $N_R=2$ one of them can have $L=4$ and the other $L=-1$, and so on.
In particular $N_R=3$, is the unique case that contains the usual solution with the three neutrinos
having the same lepton number which is identical to the generation-by-generation case. However,
there are infinite exotic solutions, say $L=(L_1,L_2,-L_1-L_2+3)$. It means that even in the
context of the model with the gauge symmetries of the SM, the addition of that sort of neutrinos
is mandatory but their number remains arbitrary, i.e., $N_R=1,2,3,\cdots$, since the constraint
equation above has always solution in the global $(B-L)$ case for any $N_R$.
We have extended this scenario when $B-L$ is gauged and contributes to the electric charge.

We have in this models that $\Delta(B-L)
\equiv-\Delta L$ and the $(\beta\beta)_{0\nu}$ occurs through the usual mechanism with massive
neutrinos. On the other hand, the proton is appropriately stabilized because there is no dimension
five operator $\overline{Q^c}Q\overline{Q^c}L$ at the tree level. The lowest dimension effective
operators, $B-L$ conserving, that contribute to its decay are dimension eight, for instance $\Lambda^{-4}\overline{Q^c}Q\overline{Q^c}L\vert\phi\vert^2$ which induces, after the spontaneous
 symmetry breaking, interactions like $\overline{Q^c}Q\overline{Q^c}L\frac{u^2}{\Lambda^4}$
that are enough suppressed  whenever $u\ll \Lambda$. A similar analysis can be made with other
 effective operators~\cite{weinberg} including those that involve right-handed sterile
neutrinos~\cite{dias}.

We have considered here the case of a local $U(1)_{B-L}$ symmetry. In the same way, it
is also possible to build models with $U(1)_X$, where $X$ denotes any of the combinations
$L_a-L_b$, $2L_a-L_b-L_c$, with $a\not =b\not=c$, for $a,b,c=e,\mu,\tau$. In these cases
right-handed neutrinos may carry non-standard values of $X$.

\end{document}